\documentstyle[12pt]{article}
\font\eightmsb=msbm10 scaled 1200
\def\bbb#1{\hbox{\eightmsb#1}}

\def\Journal#1#2#3#4#5{#5 {#1} {\bf #2} #3}
\def\CQG{\em Class. Quantum Grav.}

\def\PRD{\em Phys. Rev. D }
\def\GRG{\em Gen. Rel. Grav.}
\def\IJT{\em Int. J. Theor. Phys.}
\def\JMP{\em J. Math. Phys.}
\def\DG{\em Diff. Geom.}
\def\CMP{\em Commun. Math. Phys.}
\def\APP{\em Acta Phys. Polon.}

\def\be{\begin{equation}}
\def\ee{\end{equation}}
\def\bea{\begin{eqnarray}}
\def\eea{\end{eqnarray}}
\def\bean{\begin{eqnarray*}}
\def\eean{\end{eqnarray*}}
\def\tr{\mbox{tr}}
\def\llp{\scriptsize}
\def\espaitemps{{\cal M}}
\def\sign{\mbox{sign}}
\def\fin{\hfill \rule{2.5mm}{2.5mm}}

\newtheorem{lemas}{Lemma}

\newtheorem{teorem}{Theorem}
\newenvironment{teo}
{\begin{teorem}{\rm\hspace{-2mm}:\hspace{2mm}}}{\end{teorem}}

\newtheorem{propos}{Proposition}
\newenvironment{propo}
{\begin{propos}{\rm\hspace{-2mm}:\hspace{2mm}}}{\end{propos}}

\newtheorem{conjetu}{Conjecture}

\newtheorem{corollary}{Corollary}[teorem]

\begin{document}
\title{Segre decomposition of spacetimes}

\author{Jos\'e M. M. Senovilla\thanks
{Also at Laboratori de F\'{\i}sica Matem\`atica, Institut d'Estudis Catalans,
Catalonia.}
\, and Ra\"ul Vera\footnotemark[1]\,\\
Departament de F\'{\i}sica Fonamental, Universitat de Barcelona,\\
Diagonal 647, E-08028 Barcelona, Catalonia.}

\maketitle
\begin{abstract}
Following a recent work in which it is shown that a spacetime admitting
Lie-group actions may be disjointly decomposed into a 
a closed subset with no interior plus a dense finite
union of open sets in each of which the character and dimension of the
group orbits as well as the Petrov type are constant, the
aim of this work is to include the Segre types of the
Ricci tensor (and hence of the Einstein tensor) into the
decomposition. We also show how this type of decomposition can be carried out
for any type of property of the spacetime depending on the existence of
a continuous endomorphism.
\end{abstract}

\section{Introduction}
In a recent work \cite{halldecomp} it has been shown that
a spacetime $\espaitemps$ admitting Lie-group actions can be decomposed
into a finite disjoint union of open subsets of $\espaitemps$
in each of which the orbit type, the dimension
of the finite Lie algebra and the
Petrov type are constant, toghether with a closed subset
that has no interior.
A first theorem deals with the Petrov type, while a second one
treats the Lie-group admitted by $\espaitemps$. Then,
both theorems are combined to give raise to a more refined
decomposition.

The motivation in \cite{halldecomp} was to ascertain whether
special solutions solve the local problem `almost everywhere'.
By special solutions it was meant those exact solutions
with a prescribed form of the energy-momentum tensor
to which some other additional simplifying assumptions
involving the Petrov types and/or existence
of symmetries were imposed.
The answer is affirmative,
because the additional assumptions (Petrov and orbit types) hold
constantly into open subsets that almost cover $\espaitemps$,
leaving aside only the points belonging to the
closed subset without interior of the decomposition,
which is of zero measure.

In the same sense, and with the idea of completing the above results,
one can wonder
whether or not it is possible to find open neighbourhoods
in which the algebraic type of the energy-momentum tensor
\cite{churchill,pleb64} does not change,
and further, whether or not the prescriptions on the Segre types
\cite{segre} solve this local problem `almost everywhere'.
As in the previous cases shown in \cite{halldecomp},
an affirmative answer would assure us that choosing {\em a priori} any
algebraic type of the energy-momentum tensor for a spacetime is licit,
because given a point with a particular Segre type
there would almost everywhere be an open
neighbourhood (any connected part of which can be taken as
our spacetime) in which the algebraic type of the
energy-momentum tensor is constant\footnote{Here we have used the term
{\it a priori} because the imposition of further
specializations may restrict the existence of such
open sets.}. Thus, the main aim of the present work consists in
adding the Segre types of the Ricci (and the
energy-momentum) tensor to the decompositions
shown in \cite{halldecomp}. To this end we will
follow \cite{halldecomp} in a very close way,
but making extensive use of results and notations of
the related papers \cite{hallrendall}-\cite{bocomo}.
Even though important physical reasons
may lead us to consider only those algebraic types of the energy-momentum
tensor which satisfy the so-called energy conditions, see for instance
\cite{HE,KRAM}, no imposition of such kind will be made in the present work
mainly because this sort of assumption does not affect the final result
and can be easily added afterwards. Thus, we have preferred to keep the
mathematical generality.

The importance of these results is not only theoretical, because
spacetimes with varying Segre type do appear many times in practice.
For example, although the standard approach for finding exact solutions
of the Einstein field equations usually starts with
the prescription of the algebraic form of the energy-momentum tensor,
the resulting spacetimes can be extendible to larger ones such
that the extension must be performed through regions that cannot keep
the same algebraic type (see e.g. section 3 in \cite{G2NC}).
More importantly, many other simplifying assumptions can be used as
the starting point (such as the existence of groups of motions or
generalized symmetries, Kerr-Schild transformations, specific Petrov
types, etc.) and they may lead to completely {\it explicit}
spacetimes having {\it subregions} with an appropriate or desired
energy-momentum tensor (vacuum, perfect-fluid, electromagnetic), but such
that, on the whole spacetime, the Segre type of the energy-momentum tensor
varies from point to point. Many illustrative examples can be found in
\cite{G2NC}-\cite{CAR2}, and some considerations on an explicit case
concerning $G_2$ spacetimes will also be given here
in Section \ref{sec:conclusions}.

The plan of the paper is as follows.
Section \ref{sec:types} is devoted to a brief review
on the Segre types and related matters concerning their
characterizations. These will be used in Section \ref{sec:decomp}
in order to proof the decomposition of any spacetime
according to the algebraic type of the Ricci tensor. Finally, in Section
\ref{sec:conclusions} some conclusions and examples are presented.
In particular, we devise a completely general method (similar to that
used in Section \ref{sec:decomp}) which allow to decompose any spacetime
in a similar manner according to the different algebraic types of any
continuous endomorphism which can act on tangent vector spaces of the
spacetime. These general decompostions may be of some importance if the
given endomorphism is related to some relevant properties of mathematical
or physical interest.

The spacetime $\espaitemps$ is assumed to be of class $C^r$
with $r\geq 2$ in order to assure a continuous Ricci tensor.
If the energy-momentum tensor were not continuous, the results on this paper
would no longer hold, and the corresponding generalized results should be
obtained by using the adequate junction conditions and their consequences.
The interior of a subset $A\subset \espaitemps$ will be denoted by
$A^\circ$.

\section{Segre types}
\label{sec:types}
The algebraic classification of a second-order symmetric tensor
\cite{pleb64} like the Ricci tensor can be obtained by various
methods, and we refer to the introduction of \cite{bocomo} for a complete set
of references on this topic and other classification schemes,
and to \cite{KRAM,hall84} for reviews. We simply note in passing that
the classification of the energy-momentum tensor is equivalent to that of the
Ricci tensor via the Einstein field equations. In fact,
it is more convenient to work on the trace-free Ricci tensor, as the
Segre type for both the Ricci tensor and its trace-free
part coincide.
Considering then the traceless part of the Ricci tensor as a linear
map on a four-dimensional real vector space, its algebraic classification
consists in the resolution of the eigenvalue problem.  The different types and
their corresponding subtypes will be denoted by the usual Segre
notation \cite{segre,KRAM}.
We shall also use the notation introduced in \cite{Limits}
and \cite{bocomo} for the {\em characteristic polynomial} (CP), the
{\em minimal polynomial} (MP) and the invariants used in the determination
of the form of the CP (see \cite{bocomo}):
given the trace-free part of the Ricci tensor $N$ by
\be
N^{\alpha}_{\;\beta}\equiv R^{\alpha}_{\;\beta}-\frac{1}{4}
\delta^{\alpha}_{\;\beta}R,
\ee
the CP of $N$ is $P(\lambda)\equiv \det(N^{\alpha}_{\;\beta}-
\lambda \delta^{\alpha}_{\;\beta})$ and takes the form
\be
P(\lambda)=\lambda^4-\frac{a}{2}\lambda^2-\frac{b}{3} \lambda +
\frac{1}{4} \left(\frac{a^2}{2}-c \right),
\ee
where $a \equiv \tr N^2$, $b \equiv \tr N^3$ and $c \equiv \tr N^4$.
In general, $P(\lambda)$ factorizes as
\be
(\lambda - \lambda_1)^{d_1}(\lambda - \lambda_2)^{d_2} \cdots
(\lambda - \lambda_r)^{d_r} ,
\ee                  
where $\lambda_i \in \bbb{C}$
and $d_i\,\,\, (i=1,\ldots,r)$ are the eigenvalues and their corresponding
degeneracies respectively. The MP of $N$ is given by
the lowest-degree monic matrix polynomial in $N^{\alpha}_{\;\beta}$
vanishing identically, which is unique and can always be factorized as
\be
(N - \lambda_1 \bbb{I} )^{m_1} (N - \lambda_2 \bbb{I})^{m_2} \cdots
(N - \lambda_r \bbb{I} )^{m_r},
\ee
where $m_i (\leq d_i)\,\,\, (i=1,\ldots,r)$ are the maximal dimensions
of the Jordan matrices for each eigenvalue.
The CP is denoted by the list
of the degeneracies $\{d_1 d_2 \cdots d_r\}$, using in particular the
notation $\{z \bar{z}
\ldots \}$ when the pair of complex conjugate eigenvalues appear
(real second-order symmetric tensors in a Lorentzian spacetime
have at most a pair of complex conjugate eigenvalues \cite{pleb64}).
The MP is indicated through the list
$\| m_1 m_2 \cdots m_r \|$, and finally, the standard
Segre symbols are used to denote the Segre types as can be
seen in Table \ref{taulaSegrePCPM} (see \cite{Limits}).

Following \cite{bocomo} one can define the invariants
\bea
I_1\equiv I_3^3-\left[ 3aI_3+4(3b^2-a^3)\right]^2,\hspace{1cm}
I_2\equiv 2a-\sqrt{|I_3|},\hspace{1cm}
I_3\equiv 7a^2-12c,
\label{inv}
\eea
whose signs will determine the form of the CP \cite{bocomo},
(see also \cite{pleb64,LUD-SCHAN,jolyMac,seixas}
for other sets of invariants), according to the values
shown in Table \ref{taulaSegrePCPM}. Some sign combinations
cannot occur, and this implies, in fact, the existence
of at most a pair of complex conjugate eigenvalues.

\begin{table}
\begin{center}
\begin{tabular}{c||c|c|c|c|c|c|c||}
CP $\rightarrow$  
           &\{1111\}&\{$z\bar{z}$11\}&\{$z\bar{z}$2\}& \{211\}  & \{31\}
           & \{22\}      & \{4\}    \\
\hline
INV $\rightarrow$
           &$I_1>0$ & $I_1<0$ & $I_1=0$&$I_1=0$&$I_1=0$&$I_1=0$&$I_1=0$    \\
           &$I_2>0$ &&$I_2<0$&$I_2>0$&$I_2>0$&$I_2=0$&$I_2=0$ \\
\cline{1-1}
MP$\downarrow$&     &         &$I_3>0$&$I_3>0$&$I_3=0$&$I_3>0$&$I_3=0$   \\
\hline\hline
$\|1111\|$ & [1,111]& [$z\bar z11$]  &               &          &          &             &          \\
           &\llp{rank 4,3}&\llp{rank 4,3}&&&&& \\ \hline
$\|211\|$  &        &                &               & [211]    &          &             &          \\
           &&&&\llp{rank 4,3}&&& \\ \hline
$\|31\|$   &        &                &               &          & [31]     &             &          \\ 
           &&&&&\llp{rank 4}&& \\ \hline 
$\|111\|$  &        &                &[$z\bar z(11)$]& [1,1(11)]&          &             &          \\
           &&&\llp{rank 4,2}&\llp{rank 4,3,2}&&& \\
           &        &                &               & [(1,1)11]&          &             &          \\
           &&&&\llp{rank 4,3,2}&&& \\ \hline 
$\|21\|$   &        &                &               &          & [(21)1]  & [2(11)]     &          \\
           &&&&&\llp{rank 4}&\llp{rank 4,1}& \\ \hline
$\|3\|$    &        &                &               &          &          &             & [(31)]   \\
           &&&&&&&\llp{rank 2} \\  \hline 
$\|11\|$   &        &                &               &          & [(1,11)1]& [(1,1)(11)] &          \\
           &&&&&\llp{rank 4}&\llp{rank 4}&  \\ 
           &        &                &               &          & [1,(111)]&             &          \\ 
           &&&&&\llp{rank 4}&& \\   \hline 
$\|2\|$    &        &                &               &          &          &             & [(211)]  \\
           &&&&&&&\llp{rank 1} \\  \hline 
$\|1\|$    &        &                &               &          &          &             & [(1,111)] \\  
           &&&&&&&\llp{rank 0} \\  
\hline\hline
\end{tabular}
\end{center}
\caption[]{Segre types of the traceless part of the Ricci tensor.
The columns indicate the
different characteristic polynomial (CP) forms together with their
corresponding characterization through the signs of the invariants
(INV) introduced in (\ref{inv}), while the rows share the same
minimal polynomials (MP).}
\label{taulaSegrePCPM}
\end{table}

In what follows, the set of points of the manifold $\espaitemps$
in which the Ricci tensor
has a specific Segre type will be called
``Segre set'' and denoted by the corresponding Segre symbol.
For instance, the Segre set $[1,(111)]$ will denote the set of points
in $\espaitemps$ where the Ricci tensor has that Segre type.
Of course, all the Segre sets constitute a collection of disjoint sets
whose union covers the whole spacetime.
Furthermore, the notation for the CP will be also
used to indicate the union of all the Segre sets lying in the same column of
Table \ref{taulaSegrePCPM}. For instance, we have that
$\{22\}=[2(11)]\cup [(1,1)(11)]$.

Another important feature that we will need later
is the character of the eigenspaces associated to the simple
and the double eigenvalues of $N$. We will again use the results in
\cite{bocomo} where there is a section treating the
covariant determination of the character of these eigenspaces by means of
a simple straightforward method which will be very useful
for our purposes, and which allow us to avoid the more involved
equivalent splittings such as that in \cite{LUD-SCHAN}. Thus,
following \cite{bocomo}, the main object to deal with
the determination of the character of the eigenspaces
is the so-called {\em \/eigentensor} of $N$ corresponding
to the eigenvalue $\lambda$, denoted by $N_\lambda$,
which is merely the MP of $N$ without a $\lambda$-factor, that is to say,
the MP of $N$ at a given point can be decomposed
in terms of the $\lambda$-eigentensor $N_\lambda$ as follows :
$(N-\lambda \bbb{I})N_\lambda$.
The eigentensor for simple eigenvalues is, in fact, a projector
onto the corresponding one-dimensional eigenspace, that is
$N_\lambda (\vec{w})=\alpha^{}_{\vec{w}}\vec{v}_\lambda$, being
$\vec w$ an arbitrary vector field and $\vec{v}_\lambda$
the eigendirection corresponding to $\lambda$. In addition,
it can be shown \cite{bocomo} that
\be
\left( \vec v_\lambda \cdot \vec v_\lambda \right)=
\frac{1}{\alpha^2_{\vec{w}}}\left(\tr N_\lambda \right)
N_\lambda(\vec w,\vec w),
\label{23BCM}
\ee
whenever $\vec w$ is not orthogonal to $\vec{v}_\lambda$,
which ensures that $\alpha_{\vec{w}}\neq 0$, and
where we have used the notation
$N_\lambda(\vec w,\vec w)\equiv \left(N_\lambda(\vec w)\cdot \vec w\right)$.
When $\lambda$ is not a simple eigenvalue
it is necessary to consider the following double 2-form
constructed from the eigentensor $N_\lambda$:
$\left( {\cal N}_\lambda \right)_{\alpha\beta\mu\nu}
\equiv \left( N_\lambda \right)_{\alpha\mu}
\left( N_\lambda \right)_{\beta\nu}
- \left( N_\lambda \right)_{\alpha\nu}
\left( N_\lambda \right)_{\beta\mu}$. Let us denote by $E_\lambda$
the 2-eigenspace associated with $\lambda$, and let $\vec v$ and
$\vec w$ be two linearly independent vectors in $E_\lambda$.
Analogously to the previous case, ${\cal N}_\lambda$
projects any 2-form $F$ to $E_\lambda$ as follows
$\left( {\cal N}_\lambda\right)^{\;\;\;\;\mu\nu}_{\alpha\beta}
F^{}_{\mu\nu} =\alpha^{}_F F^{(\lambda)}_{\alpha\beta}$,
where $F^{(\lambda)}_{\alpha\beta}\propto v_{[\alpha}w_{\beta]}$.
If $\alpha^{}_F\neq 0$, it can be shown \cite{bocomo} that
\be
{\cal G}( F^{(\lambda)},F^{(\lambda)} )=
\frac{1}{\alpha^2_F} \left( \tr N_\lambda \right)^2 {\cal N_\lambda}(F,F),
\label{27BCM}
\ee
where ${\cal G}_{\alpha\beta\mu\nu}=g_{\alpha\mu}g_{\beta\nu}-
g_{\alpha\nu}g_{\beta\mu}$ is the usual bivector metric
so that ${\cal G}(F^{(\lambda)},F^{(\lambda)})$
is positive (resp. negative) when $E_\lambda$ is spacelike
(resp. timelike).

\section{The decomposition}
\label{sec:decomp}
Let us introduce first some preliminary results.
Regarding the continuity and differentiability of the
eigenvalues of  mixed second-order tensors
we have that (see e.g. \cite{hallrendall,dieudonne} for proofs): 

(i) The simple roots of a polynomial depend differentiably
on the polynomial coefficients.

Thus, the simple eigenvalues ($\lambda_i; \, \, d_i=1$)
of $N$ at a given point $p$ give rise to
differentiable functions (continuous if $N$ is only
continuous) in some open neighbourhood $V_p$ of $p$,
such that $\lambda_i(q)$ are the simple eigenvalues of $N$
at every $q\in V_p$.

(ii) If a continuous mixed second-order tensor $T$ has the same form of
its CP in an open set $U\subseteq \espaitemps$, then there is an open
neighbourhood $V_p\subset U$ for each $p\in U$ and continuous maps
$\lambda_1, \dots, \lambda_r\,:\, V_p \rightarrow \bbb{C}$
giving the eigenvalues of $T$ at each point in $V_p$.

By the `same form of the CP'  we mean
the same number of different roots and the
same set of degeneracies ($d_r$), that is, the same notation $\{ \dots \}$.
In \cite{hallrendall} statement (ii) assumed
`same algebraic type' (that is, same Segre type), but in fact
the proof given there does not need this restriction. On the other hand,
the continuity of the eigenvalues is sufficient for
what follows, so we have preferred not to demand extra 
differentiability assumptions on $N$.

The continuity of $N$ has some immediate consequences. First of all,
consider the subset $[1,111]\subseteq \espaitemps $, whose points
$p\in[1,111]$ are invariantly determined by $I_1(p)>0$ (see Table
\ref{taulaSegrePCPM}).
Since the invariants are continuous functions,
it follows that $[1,111]^\circ=[1,111]$. The same reasoning works for
$[z\bar{z}11]$, where $I_1<0$.
Thus, $[1,111]$ and $[z\bar{z}11]$ are two open subsets of $\espaitemps$. 

Another preliminary result comes from the rank theorem applied
to $N$. 
This theorem \cite{dieudonne} states that the rank of a continuous linear
map in some open neighbourhood of a point $p$ is equal or greater
than it is at $p$.
The possible ranks that $N$ can have for each Segre type
are also shown in Table \ref{taulaSegrePCPM}.
Let us define
\[
Y\equiv [1,111]\cup [z\bar{z}11] \hspace{1cm}\Longrightarrow \hspace{1cm}
Y=Y^\circ \, ,
\]
\[Z\equiv  \{z\bar{z}2\}\cup \{211\}\cup \{31\}\cup \{22\}
\cup [(31)]\cup [(211)].
\]
Since the ranks in the set 
$Y\cup Z$ are greater than zero, the rank theorem implies that
this union constitutes an open set. We are now ready to prove the
following intermediate step.

\begin{propo}
Any spacetime $\espaitemps$ can be decomposed in the following union of
disjoint subsets
\bea
&&\espaitemps=[1,111]\cup \left(\frac{}{}\![1,1(11)]\cup [(1,1)11]
\right)^\circ 
\cup [(1,1)(11)]^\circ 
\cup \left(\frac{}{}\![(1,11)1]\cup [1,(111)]\right)^\circ
\hspace{0cm}\nonumber\\
\label{prelim.decomp}
&&\hspace{1cm}\cup [(1,111)]^\circ \cup [211]^\circ
\cup [2(11)]^\circ 
\cup [(21)1]^\circ \cup [(211)]^\circ \\
&&\hspace{1cm}\cup [31]^\circ \cup [(31)]^\circ
\cup [z\bar{z}11]\cup [z\bar{z}(11)]^\circ \cup X. \nonumber
\eea
with $X^\circ=\emptyset$, where $X$ is the necessarily closed subset
defined from the decomposition.
\label{1decomp}
\end{propo}

{\em Proof:}
From the disjointness of the decomposition we know that
$X\cap [1,111] =X\cap [z\bar{z}11]=X\cap Y=\emptyset$, so that
$I_1(p)=0$ for every $p\in X$:
\[
I_1(X)=0\, .
\]
The only thing we have to prove is that the set $X$, called the
{\it remainder},
has no interior. Suppose then, on the contrary, that there existed an open
subset $W$ such that $\emptyset \neq W \subseteq X$. From the definition of the
remainder and the above results, and since the union $Y\cup Z$ is open,
we would have that $W\cap (Y\cup Z)=W\cap Z$ is open as well.
If the open subset $W\cap Z$ were non-empty, several possibilities
would appear.

Possibility 1: 
Suppose that $W\cap  \{z\bar{z}2\}=
W \cap  Z \cap  \{z\bar{z}2\}\neq \emptyset$.
Let $p$ be a point in the intersection. As in particular
$p \in \{z\bar{z}2\}$, this would mean that $I_1(p)=0, I_2(p)<0$
(see Table \ref{taulaSegrePCPM}),
and therefore, since the invariants are continuous functions,
there would exist an open neighbourhood $U$
of $p$ such that $I_2(U)<0$. Furthermore,
we could choose $U$ such that $U\subset W\cap Z$ because $p$ belongs also
to the open set $W\cap Z$. This would also imply $I_1(U)=0$. Thus, we
would have
that $U\subset W\cap \{z\bar{z}2\}$, which immediately implies
$W\cap \{z\bar{z}2\}^\circ \neq \emptyset$,
in contradiction with the disjointness in the definition of $X$.
Thus, possibility 1 is not valid and
\[
W\cap  \{z\bar{z}2\}=\emptyset.
\]

Possibility 2:
Suppose then that $W\cap \{211\}=W\cap Z\cap \{211\}\neq \emptyset$.
Following the same reasoning as above, there would be a point $p$ in this
intersection, so $p\in \{211\}\Leftrightarrow \{ I_1(p)=0,I_2(p)>0,I_3(p)>0\}$,
and therefore, there would be an open neighbourhood $U$ of $p$ such that
$I_2(U)>0,I_3(U)>0$. And again, as $p\in W \cap  Z$, we could choose
$U\subset W \cap  Z$ so that $I_1(U)=0$, and we would have
then $U\subset W\cap \{211\}$.

Now, from the definition of the remainder we have that $W\cap
([1,1(11)]\cup [(1,1)11])^\circ =\emptyset$, which
implies that $U\subset W$ cannot intersect
$\left( [1,1(11)]\cup [(1,1)11]\right) ^\circ$, and in particular
$U$ cannot be entirely contained in $[1,1(11)]\cup [(1,1)11]$.
This would mean that $U\cap [211]\neq \emptyset$.
There should then be a point $m\in U$ such that $m\in [211]$.
The fact that, in particular, $U\subset \{211\}$ would assure that
the CP of $N$ will have the same form all over $U$. Therefore, statement (ii)
could be applied to $N$ in $U$ in order to state that
for every point in $U$ we could choose an open neighbourhood $V\subset U
\subset W\cap \{211\}$ where three continuous functions
$\lambda_1, \lambda_2, \lambda_3 \,:\, V\longrightarrow \bbb{R}$
would give the values of the three different eigenvalues of $N$ in $V$.
The linear map
\[
{\cal Q}\equiv \left( N-\lambda_1 \bbb{I} \right)
\left( N-\lambda_2 \bbb{I} \right)\left( N-\lambda_3 \bbb{I} \right),
\]
would be continuous in $V$ from where we could deduce the existence of an
open neighbourhood $B\subset V$ of $m$ where ${\cal Q}(B)\neq 0$, given that
${\cal Q}(m)\neq 0$. This would in turn
imply that $m\in B \subset [211]^\circ$, and hence
$m\in W\cap [211]^\circ\Rightarrow W\cap [211]^\circ\neq \emptyset$,
in contradiction with the disjointness in the definition of $X$.
Thus, possibility 2 is not valid and
\[
W\cap \{211\}=\emptyset.
\]

Possibility 3:
Let us assume then that $p\in W\cap \{31\}=W\cap Z\cap \{31\}$.
Since $p\in \{31\} \Leftrightarrow
\{ I_1(p)=I_3(p)=0,I_2(p)>0 \}$, there would be an open neighbourhood
$U'$ of $p$ where $I_2(U')>0$ and one could take the necessarily open set
$U\equiv U'\cap \left( W\cap Z\right)$ containing $p$, which would have
$I_1(U)=0$ and $I_2(U)>0$. But this would further imply $I_3(U)=0$ because
$W\cap \{211\}=\emptyset$ (see Table \ref{taulaSegrePCPM}),
so that $U\subset W\cap \{31\}$.

As the open set $U$ cannot be entirely contained in
$\left( [(1,11)1]\cup [1,(111)] \right)^\circ$ because
$W\cap \left( [(1,11)1]\cup [1,(111)] \right)^\circ=\emptyset$,
there should be at least a point $m\in U$ such that
$m\in [31]\cup [(21)1]$. Since $U\subset \{31\}$
the linear map
$N$ would have the same form of its CP
everywhere in $U$, and thus result (ii) could be applied.
Let $\lambda_1, \lambda_2:V\longrightarrow \bbb{R}$ be the two
continuous functions defined in an open neighbourhood $V\subset U$ of $m$
giving the eigenvalues of $N$ in $V$. In $V$ we could consider the two
linear maps defined by
\bean
&&{\cal Q}_1\equiv \left( N-\lambda_1 \bbb{I} \right)^2
\left( N-\lambda_2 \bbb{I} \right), \\ 
&&{\cal Q}_2\equiv \left( N-\lambda_1 \bbb{I} \right)
\left( N-\lambda_2 \bbb{I} \right). 
\eean
If $m$ were in [31], we would have that ${\cal Q}_1(m)\neq 0 $ and
thus there would exist an open neighbourhood $B\subset V$ of $m$ such that
${\cal Q}_1(B)\neq 0 $, which would mean that $m\in B\subset [31]^\circ$
and therefore that $m\in W \cap [31]^\circ$, in contradiction with the
definition of the remainder. Hence $U\cap [31]=\emptyset$. Then, it would
necessarily follow that $m\in U\cap [(21)1]$. But a similar reasoning,
using now ${\cal Q}_2(m)\neq 0 $, would provide
$W \cap [(21)1]^\circ\neq \emptyset$,
in contradiction with the disjointness of the decomposition defining $X$.
Thus, possibility 3 cannot hold and
\[
W\cap \{31\}=\emptyset. 
\]

Possibility 4: Assume now there were a
$p\in W\cap \{22\}=W\cap Z \cap  \{22\}$.
Summing up the results up to now, we know that $I_1(W)=I_2(W)=0$, and since
$p\in \{22\}  \Leftrightarrow \{ I_1(p)=I_2(p)=0,I_3(p)>0 \}$, there would
be an open neighbourhood $U'$ of $p$ where $I_3(U')>0$. One could
then consider the necessarily open set $U\equiv U'\cap (W\cap Z)$,
so that $I_1(U)=I_2(U)=0$ and $I_3(U)>0$. This would mean
$p \in U\subset W\cap \{22\}$.

As always, the disjointness of the decomposition would imply that the
open set $U$ cannot be entirely contained in $[(1,1)(11)]$.
Therefore, there should be a point $m\in U\cap [2(11)]$.
Since in particular $U\subset \{22\}$,
we could again apply result (ii) to assure the
existence of an open neighbourhood $V\subset U$ of $m$ where the
two continuous functions 
$\lambda_1, \lambda_2\, :\,V\longrightarrow \bbb{R}$ would be the
two double eigenvalues of $N$ at each point in $V$.
Using the continuous linear map defined in $V$ and given by
\[
{\cal Q}_3 \equiv \left( N-\lambda_1 \bbb{I} \right)
\left( N-\lambda_2 \bbb{I} \right),
\]
and as ${\cal Q}_3 (m)\neq 0$, there would be an open neighbourhood
$B\subset V$ of $m$ such that ${\cal Q}_3 (B)\neq 0 \Rightarrow
B\subset [2(11)]^\circ \Rightarrow m\in W \cap [2(11)]^\circ$,
against the disjointness in the definition of $X$.
Thus, possibility 4 is, in fact, not possible and
\[
W\cap \{22\}=\emptyset.
\]

Possibility 5:
The only way to keep a non-empty $W\cap Z$ is that there were a point
$p\in W\cap \left([(31)]\cup [(211)]\right) =W\cap Z$, so that
$\{ I_1(p)=I_2(p)=0,I_3(p)=0 \}$. Hence, there would be an open
neighbourhood $U$ of $p$ such that $U\subset W\cap Z\Rightarrow
U\subset (\{4\}-[(1,111)])$, and $U$ could not be entirely contained in
[(211)] in order to keep the disjointness of the decomposition. Then, there
should be a point $m\in U\cap [(31)]$ and, as in particular
$U\subset \{4\}$, applying again (ii) one could construct
an open neighbourhood $V\subset U$ of $m$ where the continuous function 
$\lambda\, :\, V\longrightarrow \bbb{R}$ would be the
quadruple eigenvalue of $N$ and such that
\[
{\cal Q}_4\equiv \left( N-\lambda \bbb{I} \right)^2
\]
would be a continuous linear map all over $V$.
As ${\cal Q}_4(m)\neq 0$, there would be an open neighbourhood $B\subset V$
of $m$ such that ${\cal Q}_4(B)\neq 0\Rightarrow
B\subset [(31)]^\circ \Rightarrow m\in W \cap [(31)]^\circ$,
contradicting once more the definition of the remainder $X$. Thus possibility
5 cannot hold either and, in summary, the initial assumption of a non-empty
set $W\cap Z$ is not valid. Hence
\[
W\cap Z=\emptyset.
\]

Now that we know that $W\cap Z$ must be empty, the only way to have a non-empty
interior for $X$ would be that the open set $W\subseteq X$ satisfied
$W\cap [(1,111)]\neq \emptyset$. But this is impossible, because
$\espaitemps=Y\cup Z \cup [(1,111)]$ and from the previous results
we know that $W=W\cap \espaitemps=W\cap [(1,111)]$,
and therefore $W\cap [(1,111)]$ would be an open set, so that
$W\cap [(1,111)]=W\cap [(1,111)]^\circ=\emptyset$
from the definition of $X$. This finally proves that
\[
X^\circ=\emptyset \, ,
\]
that is to say, the remainder $X$ defined from the decomposition
(\ref{prelim.decomp}) has no interior in the manifold topology. \fin

\vspace{5mm}

Once that we have Proposition \ref{1decomp} at hand, and in order to achieve
the main result, we must only refine the above decomposition using the
character
of the eigenspaces of $N$ to `separate' the sets sharing the same
box in Table \ref{taulaSegrePCPM}, which are
packaged toghether in open union sets in (\ref{prelim.decomp}).
To this end, one could make use of the results given in \cite{Limits}
on limits of spacetimes applied to continuous paths
between points within a given spacetime. This would need the explanation of the
more refined algebraic classification of the Ricci tensor presented in
\cite{LUD-SCHAN} where some different invariants are used.
We refer the reader to \cite{penrose1}, \cite{cradehall} and \cite{hallhungary}
where this more refined classification is studied (and compared with the one
used herein)
by means of geometrical and topological considerations.
However, the result can be also obtained while keeping the
simplicity as follows.

\begin{teo}
Any spacetime $\espaitemps$ can be decomposed in the following union of
disjoint subsets
\bean
&&\espaitemps=[1,111]\cup [1,1(11)]^\circ \cup [(1,1)11]^\circ 
\cup [(1,1)(11)]^\circ 
\cup [(1,11)1]^\circ \cup [1,(111)]^\circ \hspace{1cm}\\
&&\hspace{1cm}\cup [(1,111)]^\circ \cup [211]^\circ
\cup [2(11)]^\circ 
\cup [(21)1]^\circ \cup [(211)]^\circ \\
&&\hspace{1cm}\cup [31]^\circ \cup [(31)]^\circ
\cup [z\bar{z}11]\cup [z\bar{z}(11)]^\circ \cup X,
\eean
with $X^\circ=\emptyset$, where $X$ is the necessarily closed subset
defined from the decomposition.
\end{teo}
Thus $\espaitemps -X$ is an open dense subset of $\espaitemps$.
Since $\espaitemps$ is connected, $X$ is empty if and only if the
Segre type is constant on $\espaitemps$.

{\em Proof:}
Given the preliminary decomposition of Proposition \ref{1decomp}, we only
need to prove the following two statements,

\begin{enumerate}
\item[(a)] $\left(\frac{}{}\![1,(111)]\cup [(1,11)1] \right)^\circ=
[1,(111)]^\circ\cup [(1,11)1]^\circ$,
\item[(b)] $\left(\frac{}{}\![1,1(11)]\cup [(1,1)11] \right)^\circ=
[1,1(11)]^\circ\cup [(1,1)11]^\circ$,
\end{enumerate}
from which the theorem follows immediately.

Consider first statement (a). Take any point
$p\in \left( [1,(111)]\cup [(1,11)1] \right)^\circ$ so that there is an
open neighbourhood $U'$ of $p$ contained in
$\left( [1,(111)]\cup [(1,11)1] \right)^\circ$ (if there is no point in
$\left( [1,(111)]\cup [(1,11)1] \right)^\circ$, then the result is trivial).
Suppose first that
$p\in [(1,11)1]\cap \left( [1,(111)]\cup [(1,11)1] \right)^\circ$.
Since $N$ has a simple eigenvalue at $p$,
and using result (i), we can choose an
open neighbourhood $U\subset U'$ of $p$ where there is 
a continuous function $\lambda :U\rightarrow \bbb{R}$ representing
the simple eigenvalue of $N$ at each point in $U$. As $N$ is trace-free,
the MP of $N$ in $U\subset \left( [1,(111)]\cup [(1,11)1] \right)^\circ$ has
the form
$\left( N-\lambda \bbb{I} \right) \left( N + \lambda \bbb{I}/3 \right)$,
and thus, the eigentensor for $\lambda$ is given by
$N_\lambda=\left( N+ \lambda \bbb{I}/3 \right)$, which hence is
also continuous all over $U$. Denoting by
$\vec v_\lambda$ the $\lambda$'s eigendirection and using 
$\tr N_\lambda=4\lambda/3$, relation (\ref{23BCM}) implies 
\be
\sign \left( \vec v_\lambda \cdot \vec v_\lambda \right)=
\sign (\lambda)\, \sign \left[ N_\lambda  \left(\vec w \cdot \vec w
\right) \right] \hspace{3mm} \mbox{in} \hspace{1mm} U,
\label{signevv}
\ee
where $\vec w$ is an arbitrary vector field that can be
chosen continuous and non-orthogonal to $\vec v_\lambda$.
At $p\in [(1,11)1]$ we have
$\sign \left( \vec v_\lambda \cdot \vec v_\lambda \right)=1$,
so this must also be the case in a sufficient small open neighbourhood
$V\subset U$ of $p$
due to the continuity of the functions involved in the righthand side
of expression (\ref{signevv}).
We have thus proven that for every
$p\in [(1,11)1]\cap\left( [1,(111)]\cup [(1,11)1] \right)^\circ$, there is an
open neighbourhood $V\subset \left( [1,(111)]\cup [(1,11)1] \right)^\circ$
such that $V\cap [1,(111)] =\emptyset$.
Then, the following chain holds
\bean
V&\subseteq &\left( [1,(111)]\cup [(1,11)1] \right)^\circ \cap V =
\left(\left( [1,(111)]\cup [(1,11)1] \right)\cap V\right)^\circ =\\
&=&\left([(1,11)1]\cap V\right)^\circ =[(1,11)1]^\circ \cap V
\eean
which means that, for every
$p\in [(1,11)1]\cap\left( [1,(111)]\cup [(1,11)1] \right)^\circ$, necessarily
$p\in [(1,11)1]^\circ$. A similar reasoning serves to prove that for any point
$p\in [1,(111)]\cap \left( [1,(111)]\cup [(1,11)1] \right)^\circ$,
there is an open neighbourhood
$V\subset \left( [1,(111)]\cup [(1,11)1] \right)^\circ$ such that
$V\cap [((1,11)1]=\emptyset$. This again implies that for every
$p\in [1,(111)]\cap\left( [1,(111)]\cup [(1,11)1] \right)^\circ$, necessarily
$p\in [1,(111)]^\circ$.
These two results together mean that for every 
$p\in \left( [1,(111)]\cup [(1,11)1] \right)^\circ$, either
$p\in [(1,11)1]^\circ$ or $p\in [1,(111)]^\circ$, from where
$\left( [1,(111)]\cup [(1,11)1] \right)^\circ=
[1,(111)]^\circ\cup [(1,11)1]^\circ$ follows.
This proves statement (a) and corresponds to the forbidden limit between
the two sets [(1,11)1] and [1,(111)] proved in \cite{Limits}.

Consider finally statement (b). Choose
any point
$p\in \left([1,1(11)]\cup [(1,1)11] \right)^\circ$ so that there is an
open neighbourhood $U'$ of $p$ contained in
$\left([1,1(11)]\cup [(1,1)11] \right)^\circ$ (again, if there is no such
point the result is immediate).
As $N$ has two simple eigenvalues at $p$, using result (i)
we can choose an open neighbourhood $U\subset U'$ of $p$ where there are
two continuous functions $\lambda_1,\lambda_2 :U\rightarrow \bbb{R}$ giving
the values of the two simple eigenvalues of $N$ all over $U$.
The eigentensor corresponding to the double eigenvalue
$\lambda_3=-(\lambda_1+\lambda_2)/2$  is given by
$N_{\lambda_3}=\left( N+ \lambda_1 \bbb{I} \right)
\left( N+ \lambda_2 \bbb{I} \right)$, which is continuous in $U$.
Since $\lambda_3$ is a simple root of the MP of $N$ in $U$,
the trace of $N_{\lambda_3}$ does not vanish in $U$, see \cite{bocomo}.
As $E_{\lambda_3}$ is spacelike at $p$, from
(\ref{27BCM}) we have that
${\cal N}_{\lambda_3}(F,F)(p)>0$ for an arbitrary
continuous simple bivector field $F$,
so ${\cal N}_{\lambda_3}(F,F)>0$ in a sufficient small
open neighbourhood $V\subset U$ of $p$. Therefore, for every
$p\in [1,1(11)]\cap\left( [1,1(11)]\cup [(1,1)11] \right)^\circ$, there is an
open neighbourhood $V\subset \left( [1,1(11)]\cup [(1,1)11] \right)^\circ$
such that $V\cap [(1,1)11] =\emptyset$. Similarly, one can prove that for every
$p\in [(1,1)11]\cap\left( [1,1(11)]\cup [(1,1)11] \right)^\circ$, there is an
open neighbourhood $V\subset \left( [1,1(11)]\cup [(1,1)11] \right)^\circ$
such that $V\cap [1,1(11)] =\emptyset$. Thus, a reasoning similar to that
used to prove statement (a) leads to
$\left( [1,1(11)]\cup [(1,1)11] \right)^\circ =[1,1(11)]^\circ \cup
[(1,1)11]^\circ$, as we wanted to show. \fin

\section{Conclusions}
\label{sec:conclusions}
The theorem just proven giving the decomposition of any spacetime
according to the Segre types of the Ricci tensor assures that,
for each point belonging to the open dense subset $\espaitemps-X$
there is an open neighbourhood in which the same Segre type holds. That is,
there are open neighbourhoods of constant Segre type almost everywhere.

This was the main aim of the paper. Nevertheless, other important 
results can arise when the theorem is applied to (or refined for) particular
cases. For example, it should be noted that,
when certain Segre types are forbidden in a given spacetime,
some additional stronger results can be obtained by just following the
characterizations via the invariants and the ranks given in Table
\ref{taulaSegrePCPM}. Besides, if the differentiability requirements
for the Ricci tensor are strengthened, for instance by demanding analyticity,
then much stronger results hold, see below for an example.
The corresponding results on limits of spacetimes
\cite{Limits} can also be very useful in the possible determination of $X$.
Of course, all of this also works for the original results \cite{halldecomp}
concerning the Petrov types.

There are several known situations in which the
presence of some regions of $\espaitemps$ with
different algebraic type of the energy-momentum tensor arise
{\em explicitly}.
Many of these situations concern spacetimes with
perfect-fluid regions admitting $G_3$ and
$G_2$ Killing groups, or three-dimensional conformal ($C_3$)
or homothetic ($H_3$) symmetry groups \cite{G2NC}-\cite{ali},
where the use of non-comoving
coordinates or merely the convenience of a system
of coordinates adapted to the symmetries implied
the imposition of the Einstein equations in such a way
that, in fact, 4 different possible Segre types
appear, namely $[1,(111)]$, $[(1,11)1]$, $[(1,111)]$,
and $[(211)]$. Sometimes this situation is hidden under
a choice of coordinates in which the solutions are actually extendible
(coordinates such as those which are valid only in the region [1,(111)]),
and such that if any extension is to be performed, it must include zones
with a different algebraic type of the energy-momentum
tensor, see \cite{G2NC} for a brief discussion.
Since in these cases there are only four possible Segre types,
further results can be shown. As an illustrative example, here we present a
result which appeared in \cite{G2NC}:

{\it 
If $\espaitemps=[1,(111)]\cup [(1,11)1] \cup [(1,111)]\cup [(211)]$,
then $[1,(111)]$ and $[(1,11)1]$ are both open sets, and hence
$[(1,111)]\cup [(211)]$ is closed.}

To see this, note first from the invariants in Table \ref{taulaSegrePCPM} that
$I_1=I_3=0$ everywhere in $\espaitemps$, hence
$[1,(111)]\cup [(1,11)1]$ is open because $I_2>0$ only there.
Taking into account the statement (a) appearing in the proof of the
main theorem, that is
$([1,(111)]\cup [(1,11)1])^\circ=[1,(111)]^\circ\cup [(1,11)1]^\circ$,
and since $[1,(111)]\cap [(1,11)1]=\emptyset$, the result follows.
In fact, from the intermediate results in the proof of (a),
which correspond to the forbidden \cite{Limits} limit between
$[1,(111)]$ and $[(1,11)1]$, it follows that
between these two sets there must always be points of $[(1,111)]\cup [(211)]$,
so it is a kind of a border.
Furthermore, if we demand analiticity for the Ricci tensor
it follows that $[(1,111)]\cup [(211)]$ has no interior, as otherwise
we would have $I_2=0$ on the whole manifold.
Finally, the remainder $X$ in the
general decomposition is, in this analytical case, the set
$[(1,111)]\cup [(211)]$ itself.

In \cite{halldecomp} the two theorems of decompositions,
the first devoted to the Petrov types (6 open subsets), and the second
to the character of the symmetry orbits (3 open subsets),
are combined to give
a third theorem in which any spacetime is decomposed into
18 disjoint open subsets in which the character of the orbit
and the Petrov type remain constants, toghether with a closed
subset without interior.
Following the same procedure, the Segre decomposition (15 open
subsets) can be combined
with the Petrov one, with the type of orbits, or with both of them,
providing then decompositions of $\espaitemps$ into
(6x15=)90, (3x15=)45, (6x3x15=)270 disjoint open subsets
plus a closed subset without interior, {\it provided} that
the combinations of Segre, Petrov and orbit types
are consistent: in many cases, the existence of isotropy
implies the degeneration of the Petrov types \cite{KRAM,EhleKundt}
and of the Ricci tensor \cite{KRAM}. For instance, the action of the
group $G_3$ on $S_2$ implies Petrov types D or O, and there
are also two equal real eigenvalues in the Ricci tensor \cite{KRAM}.
See references \cite{plebstachel,goennstach,goenner1} and the tables at the
end of the book \cite{KRAM} for studies on the relation between the existence
of isotropy groups, the algebraic classification of the Ricci 
and the Weyl tensors, and related matters. 

An important point to be raised is that, both in the results of
\cite{halldecomp} as well as in this paper, the basic ingredient is the
continuity of the maps defined by the Weyl or the Ricci tensors. Thus, it
seems feasible that similar decomposition results could be found using
other tensors which can be considered
as continuous endomorphisms on arbitrarily
dimensional vector spaces `tangent' to an $n-$dimesional manifold,
by just taking into account the continuity of both the
eigenvalues and their corresponding eigenspaces, together with
the invariants characterizing the characteristic polynomials
and also the possible ranks for each algebraic case. In what follows we present
the sketch of the {\it general} proof for any such case.

This general proof needs, first of all, an adequate table such as that used
in this paper, where the invariants discriminating between different CP and
MP are shown. Then, as a preliminary result, the spacetime is decomposed into
the union of the interior of the sets having the same CP and the same MP,
which we are going to call CPMP sets, plus the remainder $X$.
The refination of this preliminary decomposition can be left to the
end. The first aim is then to show that $X$ has no interior. To that end,
one starts by discarding the CPMP sets which are open in general
and need no further treatment. We call this set $Y$ and trivially
$X\cap Y=\emptyset$. After this, one constructs the set $Z$ defined by the
complementary of $Y$ minus the CPMP set $[(1\ldots 1)]$.
Obviously, the union $Y\cup Z$ has positive rank and is open thanks
to the rank theorem and the continuity assumed on the traceless tensor which
provides the decomposition. Moreover, $X^\circ \cap (Y\cup Z)=X^\circ \cap Z$
is open. The next step is to prove that, in fact, $X^\circ \cap Z=\emptyset$.
To do that, one takes the groups of sets with the same form of the
CP (columns),
and orders them according to the (non-strictly) increasing number of vanishing
invariants. This order is important and one has to proceed one by one
accordingly by proving that $X^\circ$ has vanishing intersection with all
of them.

Thus, one takes the `first' column (minimum number of vanishing invariants)
which will be formed by $k$ (say) CPMP sets. One can further order these $k$
sets by means of the increasing exponents of their corresponding MP, denoting
them by $\mbox{CPMP}_i$ with $i=1\dots k$, so that $\mbox{CPMP}_1$ is the
set with the minimum values of the exponents of the MP within this column.
Due to the continuity of the invariants, one can always manage to find an open
neighbourhood $U$ contained in the column, and also included in $X^\circ$
{\it whenever} the intersection of $X^\circ$ with the column is non-empty.
Then, it is possible to define continuous functions giving the eigenvalues of
our tensor inside $U$. From the disjointness of the decomposition, $U$ cannot
be entirely contained in any of the $\mbox{CPMP}_i$ sets if it is in $X^\circ$,
and in particular it cannot be included in the set $\mbox{CPMP}_1$. But then
there should be at least one point $p\in U$ lying in the union of the other
$(k-1)$ sets $\mbox{CPMP}_i$ with $i=2\dots k$. If $p$ were in the set
$\mbox{CPMP}_k$, the MP corresponding to the contiguous set $\mbox{CPMP}_{k-1}$
would not vanish at $p$ and by continuity it would not vanish in an open
neighbourhood $V\subset U$ of $p$, which would be included in both
$X^\circ$ and
$(\mbox{CPMP}_k)^\circ$, contradicting the definition of the remainder.
One repeats this reasoning orderly with the sets
$\mbox{CPMP}_{k-1},\dots, \mbox{CPMP}_2$ and finally reaches the conclusion
that the only possibility is that $X^\circ$ has no intersection with this
column. Once the `first' column has been discarded, the same procedure is used
with the next ones in order, proving that $X^\circ \cap Z=\emptyset$.
One must finally take into account that the last column was not complete
because the set with the vanishing tensor is not included in $Z$. But the
intersection of $X^\circ$ with this last set is necessarily open due to
the previous results, and therefore the disjointness of the decomposition
implies finally that this intersection is empty, that is to say, $X$ has
no interior.

The final step is the refinement separating the interiors of the various
CPMP subsets into the interiors of more specialized algebraic subsets
differing only on the character of the eigenspaces. The continuity of these
characters within a given CPMP set may be invoked for this purpose using
analogous arguments to those shown here for the proof of the main theorem.
Notice that the continuity of the character of the eigenspaces
could be proven, in principle, following intrinsic characterizations
analogous to those in \cite{bocomo}, by using the corresponding eigentensors
to construct projectors onto the different eigenspaces. Of course, other
alternative procedures, such as those in \cite{Limits}, can also be used for
this last step.

\section*{Acknowledgements}
The authors wish to thank G. S. Hall for a
fruitful correspondence and his valuable
comments and suggestions.
We also thank B. Coll for a helpful discussion and G. S. Hall, 
M. Mars and A. Molina for providing
copies of some of the references appearing in this work.
R. V. acknowledges the financial support of the {\it Direcci\'o
General de Recerca, Generalitat de Catalunya}.

\end{document}